\documentclass[aps,preprint]{revtex4}
\usepackage[dvips]{graphicx}
\begin{document}
\draft
\title{\bf Departure of some parameter-dependent spectral statistics of irregular quantum graphs from Random Matrix Theory predictions}
\author{Oleh Hul$\,^1$, Petr \v Seba$\,^{2,3}$, and Leszek Sirko$\,^1$}

\address{$^1$ Institute of Physics, Polish Academy of Sciences,
 Aleja  Lotnik\'{o}w 32/46, 02-668 Warsaw, Poland\\
$^2$ University of Hradec Kralove, V\'ita Nejed\'eho 573, 50002
Hradec Kr\'alov\'e, Czech Republic\\
$^3$ Institute of Physics, Academy of Sciences of the Czech
Republic, Cukrovarnicka 10, CZ-162 53 Praha 6, Czech Republic}

\date{May 11, 2009}

\bigskip

\begin{abstract}

Parameter-dependent statistical properties of spectra of totally
connected irregular quantum graphs with Neumann boundary
conditions are studied. The autocorrelation functions of level
velocities $c(\mathsf{x})$ and $\tilde{c}(\omega,\mathsf{x})$ as
well as the distributions of level curvatures and avoided crossing
gaps are calculated. The numerical results are compared with the
predictions of Random Matrix Theory (RMT) for Gaussian Orthogonal
Ensemble (GOE) and for coupled GOE matrices. The application of
coupled GOE matrices was justified by studying localization
phenomena in graphs' wave functions $\Psi(x)$ using the Inverse
Participation Ratio (IPR) and the amplitude distribution
$P(\Psi(x))$.

\end{abstract}


\bigskip
\maketitle

\smallskip

Hamiltonians $H(X)$ of many chaotic systems depend on external
parameters $X$. Depending on the system there is a big variety of
parameters $X$. It may be the potential
energy~\cite{SavytskyyPRERS}, the magnetic
field~\cite{SimonsPRLhyd,SimonsPRLuni,SzaferParam,SimonsPRBuni,FaasPRBuni,BerryJPAparam,BruusParam},
the shape of billiard's
boundary~\cite{SavytskyyPRERS,Hluszczuk_RS,BruusCorrel} or the
lengths of the bonds of quantum graphs~\cite{KottosAP}. For
parameter dependent systems the energy levels are determined
through the eigenvalue problem $H(X)\phi_i(X)=E_{i}(X)\phi_i(X)$.
It is conjectured that for quantum chaotic systems, when the
variation of the parameter $X$ does not change the symmetry of the
system, statistical properties of the level dynamics should be
universal
~\cite{SimonsPRLhyd,SimonsPRLuni,SimonsPRBuni,FaasPRBuni,GaspardCurv,SmolyarenkoParam}.
In order to investigate universalities in the level dynamics
several parametric statistics can be used, for example,
autocorrelation functions of the level
velocities~\cite{SimonsPRBuni}, the level curvature distribution
and the avoided crossing distribution~\cite{ZakrzewskiKusAcross}.
Parametric dynamics of random matrices was also
investigated~\cite{ZakrzewskiCurv,ZakrzewskiAcross,KunstmanNonun,GuarnieriCorrel,DelandeJPSJapan,ZakrzewskiZPB}.

Quantum graphs have attracted much attention in recent
years~\cite{KottosAP, BerkolaikoCMP, FullingJPA, GnutzmannAP,
PakonskiTanner, Exner07, BarraGaspard}. A lot of properties of
quantum graphs have already been studied. For example, spectral
properties of graphs were studied in the series of papers by
Kottos and Smilansky~\cite{KottosSmilanskyPRL}, where the authors
showed that quantum graphs are excellent paradigms of quantum
chaos. The autocorrelation functions of level velocities for
graphs with five vertices with and without time reversal symmetry
were studied in the paper~\cite{CorrelGraphs5}. However, many
other properties of graphs require much thorough consideration. In
this paper we study the autocorrelation functions of level
velocities as well as the distributions of level curvatures and
avoided crossing gaps for quantum graphs.

Graphs can be considered as idealizations of physical networks in
the limit where the widths of the wires are much smaller than
their lengths. Among the systems modelled by graphs one can find,
e.g., electromagnetic optical waveguides~\cite{Flecia87, Mitra71},
quantum wires~\cite{Ivchenko98, Gil98},  mesoscopic
systems~\cite{Imry96, Kowal90}, or microwave networks
\cite{Hul2004, Lawniczak2008}.

A quantum graph (network) consists of $n$ vertices connected by
bonds. On each bond of the graph one-dimensional Schr\"odinger
equation is defined $(\textrm{we assume that} \; \hbar=2m=1)$:
\begin{equation}
\label{Schrodinger Graf} {- \frac{d^{2}}{dx^{2}}
}\Psi_{i,j}(x)=k^{2}\Psi_{i,j}(x) \textrm{,}
\end{equation}
where $k$ is the wave vector and the subscripts $i,j$ denote the
bond which connects two vertices with the numbers $i$ and $j$.
More detailed description of quantum graphs can be found in the
paper~\cite{KottosAP}.

In order to investigate the autocorrelation functions of level
velocities one should unfold both the energies $E_{i}=k_{i}^2$ and
the parameter $X$. Energy levels of quantum graphs are unfolded by
using the mapping
\begin{equation}
\xi_{i}=N_{av}(E_{i}) \mbox{,}
\end{equation}
where $N_{av}(E)$ is the average number of states, which assures
that the mean level spacing is equal to unity. The average number
of states $N_{av}(E)$ is given by the
formula~\cite{KottosSmilanskyPRL}
\begin{equation}
N_{av}(E)=\frac{\sqrt{E}L}{\pi}+\frac{1}{2} \mbox{,}
\end{equation}
where $L$ is the total length of the graph.

The parameter $X$ was unfolded using the generalized conductance
$C_0(X)$~\cite{SimonsPRLhyd,SimonsPRLuni}:
\begin{equation}
C_0(X)=\Biggl< \biggl(\frac{\partial \xi_{i}(X)}{\partial
X}\biggr)^{2}\Biggr>\mbox{,}
\end{equation}
where $\langle\ldots\rangle$ means the average over the energy
levels. The unfolded parameter is given by the following
formula~\cite{Lebouf}:
\begin{equation}
\label{ParameterScal}
\mathsf{x}=\int\limits_{X_{in}}^{X}\sqrt{C_0(X)}dX\mbox{,}
\end{equation}
where $[X_{in},X]$ is the interval of integration. After unfolding
of the parameter $X$ statistical properties of energy levels
$\xi_{i}(\mathsf{x})$ of chaotic systems should be
universal~\cite{SimonsPRLuni,Lebouf}. Particulary, for quantum
chaotic systems the parametric level velocities distribution
$\mathsf{v}_{i}=\partial \xi_{i}/\partial \mathsf{x}$ should be
described by a Gaussian distribution~\cite{FaasPRBuni,Lebouf}.

The autocorrelation function of level velocities $c(\mathsf{x})$
is defined as follows:
\begin{equation}
c(\mathsf{x})=\biggl<\frac{\partial \xi_{i}}{\partial
\bar{\mathsf{x}}}(\bar{\mathsf{x}})\frac{\partial
\xi_{i}}{\partial
\bar{\mathsf{x}}}(\bar{\mathsf{x}}+\mathsf{x})\biggr>\mbox{,}
\end{equation}
where the average is performed over the parameter
$\bar{\mathsf{x}}$ and over all the levels. This function is the
measure of correlation of level velocities which belong to the
same energy level.

Another autocorrelation function investigated in this paper is
$\tilde{c}(\omega,\mathsf{x})$~\cite{SimonsPRBuni}
\begin{equation}
\label{Correlation}
\tilde{c}(\omega,\mathsf{x})=\frac{\sum_{i,j}\bigl<
\delta\bigl(\xi_{i}(\bar{\mathsf{x}})-\xi_{j}(\bar{\mathsf{x}}+\mathsf{x})-\omega\bigr)\frac{\partial
\xi_{i}}{\partial
\bar{\mathsf{x}}}(\bar{\mathsf{x}})\frac{\partial
\xi_{j}}{\partial
\bar{\mathsf{x}}}(\bar{\mathsf{x}}+\mathsf{x})\bigr>}{\sum_{i,j}\bigl<
\delta\bigl(\xi_{i}(\bar{\mathsf{x}})-\xi_{j}(\bar{\mathsf{x}}+\mathsf{x})-\omega\bigr)\bigr>}\mbox{,}
\end{equation}
where the average is performed over the parameter
$\bar{\mathsf{x}}$ only. This function measures the correlation of
level velocities for energy levels separated by a distance
$\mathsf{x}$ in space and by a distance $\omega$ in the unfolded
energy.

We study the autocorrelation functions of the level velocities
$c(\mathsf{x})$ and $\tilde{c}(\omega,\mathsf{x})$ for quantum
graphs. We consider only totally connected quantum graphs (every
vertex is connected with the others) with no loops and no multiple
bonds (every two vertices are connected by one bond only). The
number of vertices $n$ defines a graph size. For graphs
considered in this work it was varied between $n=4$ and $n=30$. We
impose the Neumann boundary conditions on each graph's vertex which
imply continuity of the wave function and the probability current
conservation on the vertices. $n$ vertices of a totally connected
quantum graph are connected by $B=n(n-1)/2$ bonds which gives
$B=6$ bonds for the graph with $n=4$ vertices and $B=435$ bonds for
the graph with $n=30$ vertices.

The change of the bonds lengths of a graph was chosen to be an
external parameter $X$. For the graphs with the even number of
bonds the lengths of all the bonds were changed, while for the
graphs with the odd number of bonds we changed the lengths of all
the bonds except the arbitrary chosen one.  The lengths of one
half of the arbitrary chosen bonds of a graph were increased
$L_{i,j}(X)=L_{i,j}+X$ while the lengths of the other half of the
bonds  were decreased $L_{i',j'}(X)=L_{i',j'}-X$. In this way the
total length of the graph $L=\sum_{i<j}L_{i,j}$ and the mean
density of states were independent on the parameter $X$.  The
lengths of the graph's bonds were changed in $100$ equally spaced
steps $dX$: $X=m dX$ ($m=1,...,100$).

The lengths of the bonds $L_{i,j}$ of each graph were chosen
according to the formula
\begin{equation}
L_{i,j}=\frac{L'_{i,j}}{\sum\limits_{i<j}L'_{i,j}} \mbox{,}
\end{equation}
where $L'_{i,j}$ were uniformly distributed on $(0,1]$. The total
length of the graph was equal to one $L=1$. For every size of a
graph we realized $99$ graph's configurations with different bonds
lengths. To ensure the bonds lengths to be much larger than the
wavelength $\lambda=2\pi/k$, we imposed the condition
$L'_{i,j}(X)>10\lambda$ to be satisfied for all graphs
configurations.

For each configuration of the graph about $80$ eigenvalues with
the level numbers between $822902$ and $822985$ were calculated.

\begin{figure}[!]
\begin{center}
\rotatebox{270} {\includegraphics[width=0.5\textwidth,
height=0.6\textheight, keepaspectratio]{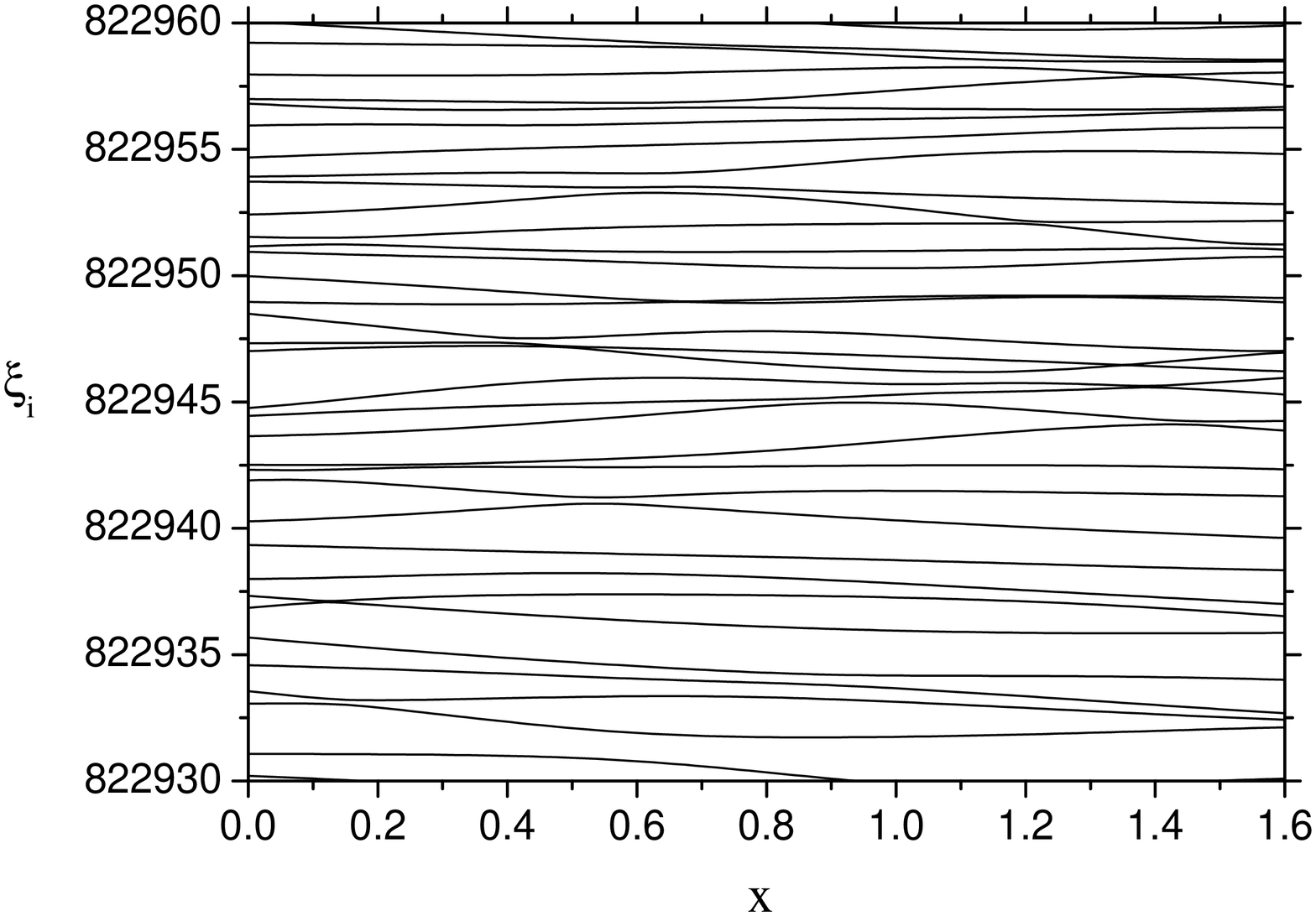}} \caption{Parametric
energy level dynamics for totally connected quantum graph with
$n=30$ vertices.} \label{Fig1}
\end{center}
\end{figure}

Figure~\ref{Fig1} shows a typical behavior of unfolded energy
levels $822930 \leq \xi_i \leq 822960$ as a function of the
external parameter $x$ for graph with $n=30$ vertices. The
parameter $x$ was unfolded using the
formula~(\ref{ParameterScal}). In Figure~\ref{Fig1} one can see
very complicated level dynamics with many avoided crossings.

Parametric level velocities of quantum graphs were calculated
using the finite difference method
\begin{equation}
\mathsf{v}_{i}=\frac{\partial \xi_{i}}{\partial \mathsf{x}}\simeq
\frac{\xi_{i}(\mathsf{x}+d\mathsf{x})-\xi_{i}(\mathsf{x})}{d\mathsf{x}}\mbox{.}
\end{equation}
The level velocities were rescaled using the variance
$\sigma_{\mathsf{v}}=\langle (\partial \xi_{i}/\partial
\mathsf{x})^{2} \rangle $
\begin{equation}
v_{i}=\frac{\partial \xi_{i}/\partial
\mathsf{x}}{\sqrt{\sigma_{\mathsf{v}}}}\mbox{.}
\end{equation}

\begin{figure}[!]
\begin{center}
\rotatebox{270} {\includegraphics[width=0.5\textwidth,
height=0.6\textheight, keepaspectratio]{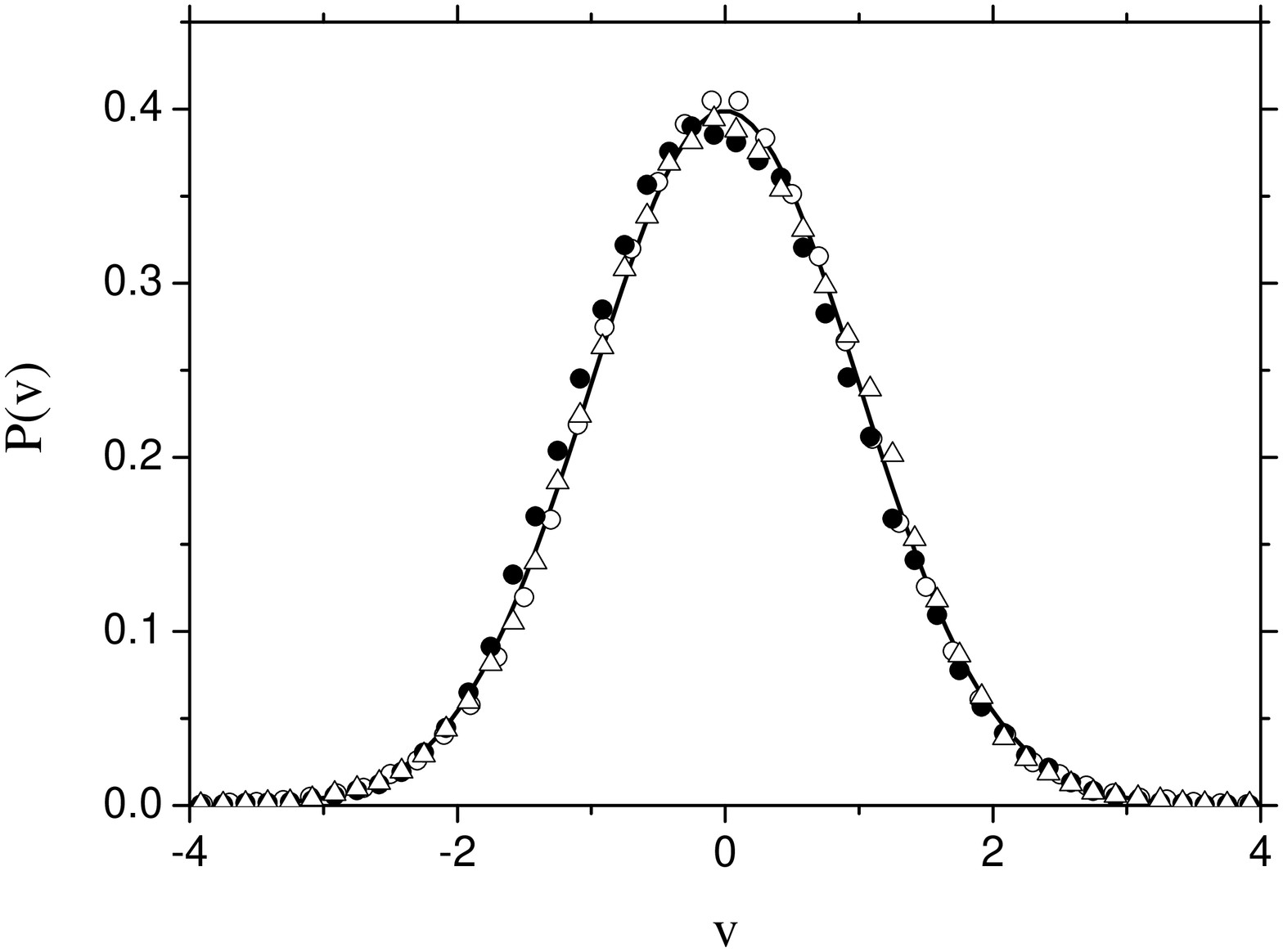}} \caption{The
parametric velocity distributions $P(v)$ for quantum graphs with
$n=6$ (open circles), $n=10$ (open triangles) and $n=30$ (full
circles) vertices, respectively. The numerical results are
compared to the Gaussian distribution (solid line).} \label{Fig2}
\end{center}
\end{figure}

Figure~\ref{Fig2} shows the velocity distribution $P(v)$ for
graphs with $n=6, 10$ and $30$ vertices. The results are averaged
over $99$ graphs configurations (approximately $8000$ data points
for each configuration). In all the cases presented in
Figure~\ref{Fig2} the velocity distribution $P(v)$ calculated for
quantum graphs is in good agreement with the Gaussian distribution
(solid line), as it is expected for quantum chaotic systems. For
the other graphs which are not presented in Figure~\ref{Fig2} with
the number of vertices between $n=6$ and $n=30$ good agreement
with the Gaussian distribution is also observed. However, for the
graphs with $n=4$ and $n=5$ vertices situation is different
(results are not shown). While for the graphs with $n=5$ the
agreement with the Gaussian distribution  is still quite
satisfactory, for the graphs with $n=4$ the velocity distribution
is lower than the Gaussian curve for small values of level
velocities. This result is not surprising, the graph with $n=4$ is
the simplest possible non-trivial fully connected graph one can
construct and for some spectral statistics it could be too simple
to exhibit behavior that is typical for quantum graphs. But when
the number of vertices of a graph increases the agreement with the
Gaussian curve improves. For this reason in our further
investigation we will concentrate mainly on graphs with $n\geq 6$
vertices.

\begin{figure}[!]
\begin{center}
\rotatebox{270} {\includegraphics[width=0.5\textwidth,
height=0.6\textheight, keepaspectratio]{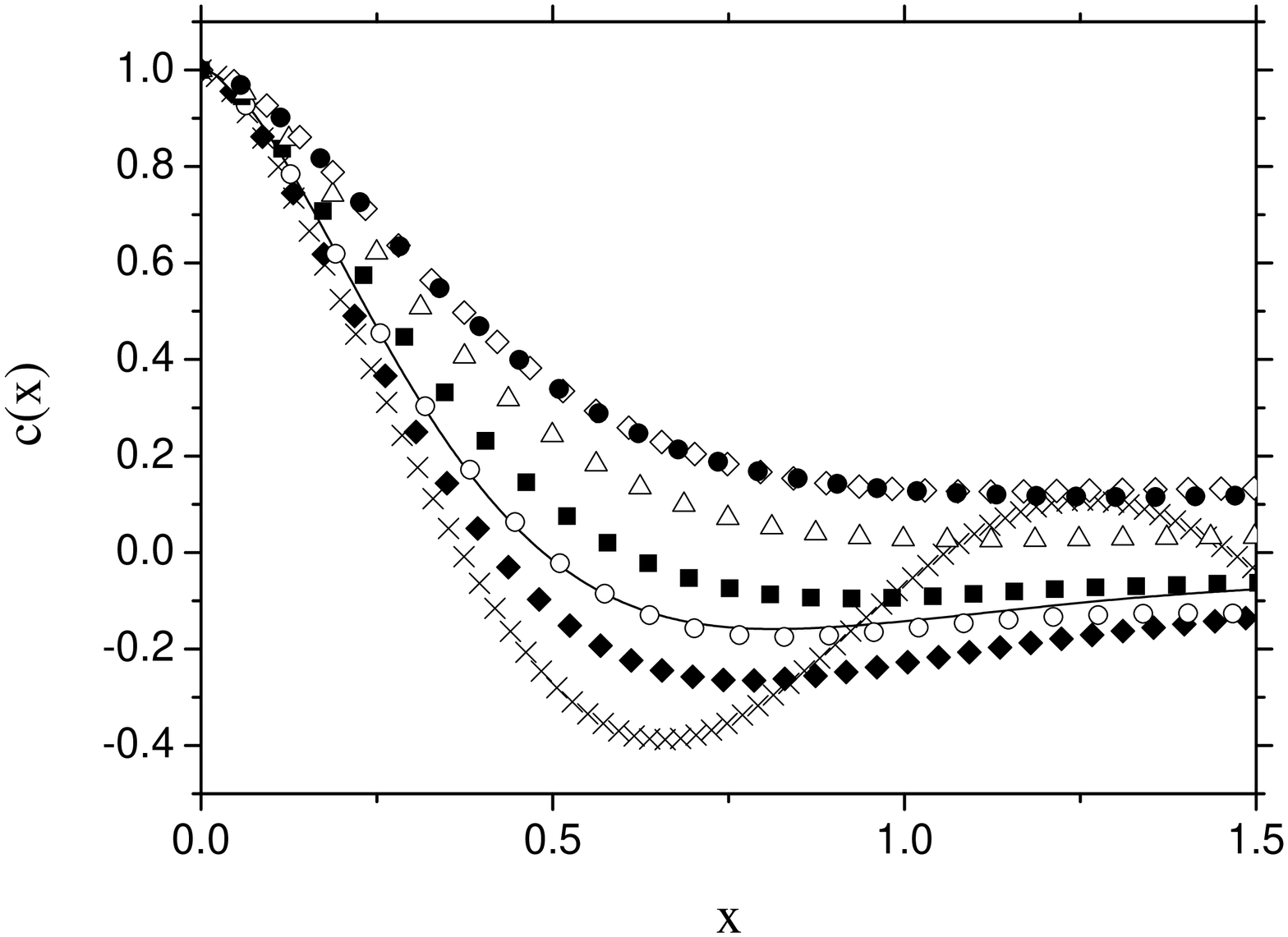}} \caption{The level
velocity autocorrelation function $c(x)$ for quantum graphs with
$n=4$ (crosses), $n=5$ (full diamonds), $n=6$ (open circles),
$n=7$ (full squares), $n=10$ (open triangles), $n=20$ (open
diamonds) and $n=30$ (full circles) compared to the predictions of
RMT for GOE (solid line).} \label{Fig3}
\end{center}
\end{figure}

Figure~\ref{Fig3} shows the autocorrelation function
$c(\mathsf{x})$ calculated for quantum graphs with
$n=4,5,6,7,10,20$ and $30$ vertices. In all cases the
autocorrelation function $c(\mathsf{x})$ was obtained by averaging
over 99 graphs configurations. Numerical results are compared with
the results of RMT for GOE (solid line). The numerical simulations
of the parameter-dependent autocorrelation function
$c(\mathsf{x})$ within RMT for GOE were made using the following
Hamiltonian model~\cite{BruusCorrel}
\begin{equation}
\label{ParametricHAmiltonian} H(X)=H_{1}\sin{(X)}+H_{2}\cos{(X)}
\mbox{,}
\end{equation}
where $H_{1}$, $H_{2}$ are two $500\times 500$ matrices that are
members of GOE. In our calculations the parameter $X$ was chosen
in 1001 equally spaced points in the interval $[0,\pi/8]$. For
each value of the parameter $X$ we ran 99 realizations of $H_{1}$
and $H_{2}$. The eigenvalues obtained from the diagonalization of
the Hamiltonian $H(X)$ were unfolded by using the integrated
average eigenvalue density for GOE matrices~\cite{Bohigas et al}.
The parameter $X$ was unfolded using the formulas (4) and (5).

As it is shown in Figure~\ref{Fig3} the autocorrelation function
of level velocities $c(\mathsf{x})$ calculated for the graphs with
$n=5$ deviates in the downward direction from the RMT predictions
for GOE (solid line) for most of the values of the parameter
$\mathsf{x}$, while for the graphs with $n\geq 7$ the
autocorrelation $c(\mathsf{x})$ deviates in the upward direction
from the RMT. The larger graph is, the bigger is deviation from
RMT. Only for the graphs with $n=6$ the agreement with RMT is
quite good. For the graphs with $n=4$ a strong deviation from RMT
is observed for $0.75<\mathsf{x}<1.5$, what is the evidence of a
non-universal behavior of the system. It is worth  mentioning that
the difference between the results obtained for big graphs with
$n=20$ and $n=30$ is very small what might suggest that for graphs
with $n>30$ the autocorrelation function $c(\mathsf{x})$ converges
to the limiting curve which is not given by the RMT predictions
for GOE.

The reason of the disagreement of $c(\mathsf{x})$ from the RMT
predictions is not exactly known. However, one can assume
\cite{SchanzScars,Kaplan01} that it is connected with non-ergodic
structures of graphs' wave functions. In order to check this
assumption we calculated the Inverse Participation Ratio (IPR)
that is a measure of wave functions localization.

\begin{figure}[!]
\begin{center}
\rotatebox{270} {\includegraphics[width=0.5\textwidth,
height=0.6\textheight, keepaspectratio]{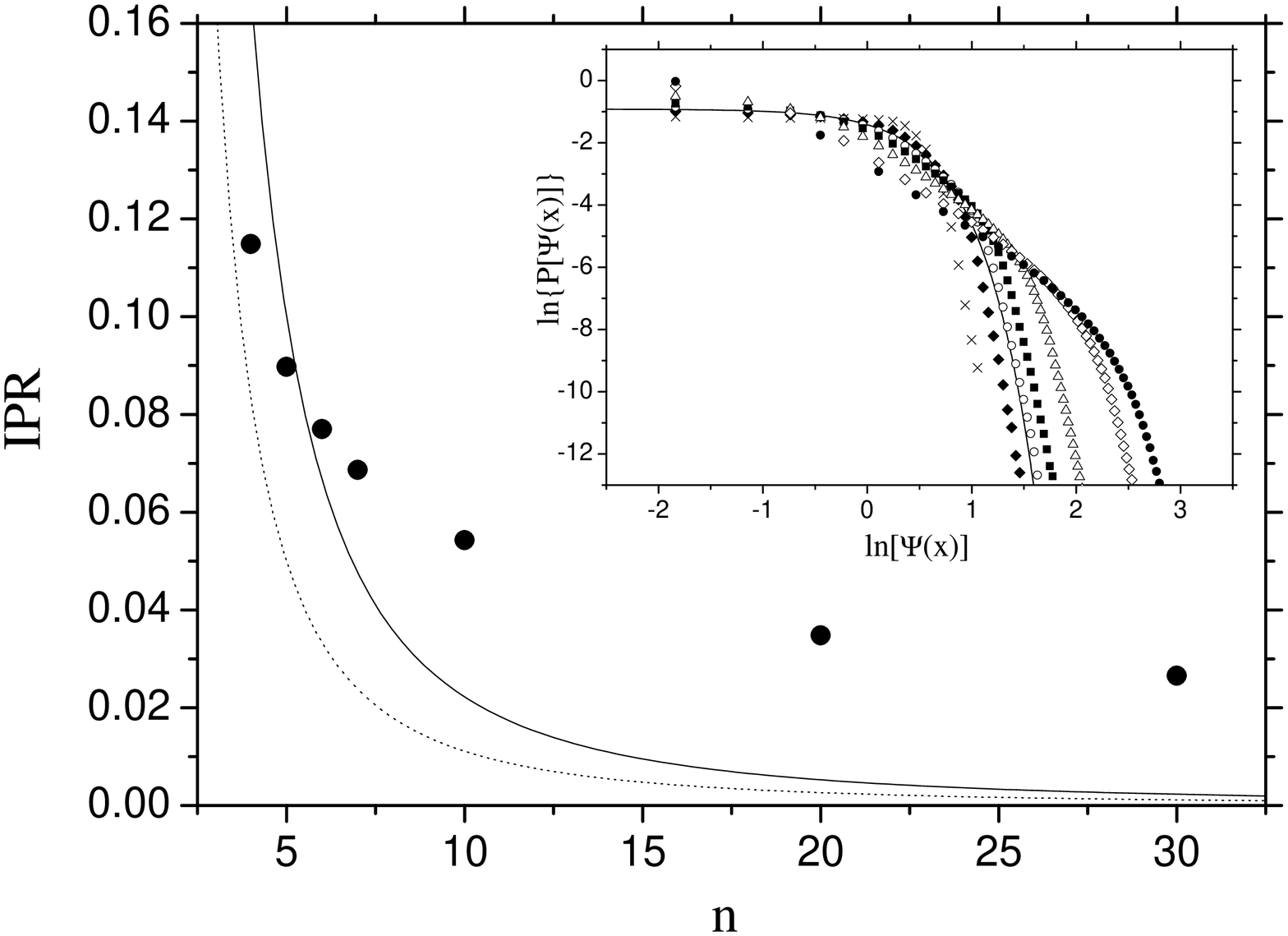}} \caption{The
inverse participation ratio (IPR) calculated for quantum graphs
(full circles) compared to the results of random wave hypothesis
(solid line). The dotted line shows the minimum values
($I_{min}=1/2B$) of IPR. In the inset we show the distributions
$P(\Psi(x))$ of the wave functions $\Psi(x)$ for quantum graphs
with $n=4$ (crosses), $n=5$ (full diamonds), $n=6$ (open circles),
$n=7$ (full squares), $n=10$ (open triangles), $n=20$ (open
diamonds) and $n=30$ (full circles) compared to the Gaussian
distribution (solid line). Distributions are presented in double
logarithmic scale.} \label{Fig4}
\end{center}
\end{figure}

In each bond of a graph a wave function can be written as
\cite{Kaplan01}:
\begin{equation}
\Psi^{(N)}_{i,j}(x)=a^{(N)}_{i,j}e^{ik_Nx}+a^{(N)*}_{i,j}e^{-ik_Nx}.
\end{equation}
Then the inverse participation ratio for the N-th energy level can
be defined in the following way:
\begin{equation}
I_N=\frac{\sum_{2B}|a^{(N)}_{i,j}|^4}{\Bigl[\sum_{2B}|a^{(N)}_{i,j}|^2\Bigr]^2}.
\end{equation}
It takes the values between $I_{min}=1/2B$ for states which occupy
each directed bond with the same probability and $I_{max}=1$ for a
state which is restricted to a single bond only, i.e., the
greatest possible degree of localization \cite{SchanzScars}.
Random Wave Hypothesis (RWH) \cite{Kaplan01} assumes Gaussian
random fluctuations of the complex coefficients $a_{i,j}$. Thus
$\langle|a_{i,j}|^4\rangle=2$ and $\langle|a_{i,j}|^2\rangle=1$.
In this case $I_{RWH}\approx 1/B$. In Figure~\ref{Fig4} we compare
the inverse participation ratio $I_n=\langle I_N \rangle$
calculated for quantum graphs of different size (circles) with the
RWH one (solid line). The averaging in the last formula was
performed over the energy levels and over graphs configurations.
The dotted line in Figure~\ref{Fig4} shows the minimum value
($I_{min}=1/2B$) of IPR. Figure~\ref{Fig4} shows that for graphs
with $n>5$ the obtained results are bigger then  RWH predictions
$I_{RWH}$. It means that wave functions of quantum graphs are less
ergodic than the ones predicted by Random Wave Model. Furthermore,
it is important to noting that the result obtained for the graphs
with $n=5$ and $n=6$ vertices are the closest ones to the
$I_{RWH}$ predictions.

We also calculated the distributions $P(\Psi(x))$ of the values of
the graph wave function $\Psi(x)$ \cite{Hlushchuk2001}. The
distributions $P(\Psi(x))$ for graphs with $n=4,5,6,10,20$ and
$30$, in the double logarithmic scale, are presented in the inset
in the Figure~\ref{Fig4}.  The distribution $P(\Psi(x))$ is
symmetric, therefore, we considered only positive values of the
wave function $\Psi(x)\geq 0$. The inset shows that the
distribution $P(\Psi(x))$ obtained for the graphs with $n=6$ (open
circles) is the closest one to the Gaussian distribution (solid
line) which is the prediction of RMT. Thus, the behavior of $I_n$
and $P(\Psi(x))$ for graphs with $n \ge 6$ suggest that the
counterintuitive results obtained for the autocorrelation function
$c(\mathsf{x})$, where the best agreement with RMT was obtained
for the graphs with six vertices, are connected with the
localization effects.

The above results suggest that in order to describe
 numerically the behavior of the autocorrelation function
$c(\mathsf{x})$ one can use the concept of coupled M-GOE matrices
\cite{Alt1998}. A coupled  M-GOE matrix with $M \times M$ blocks
can be constructed using a random GOE matrix. The values of
diagonal blocks elements of the random block matrix are equal to
the values of random GOE matrix elements, while the elements of
off-diagonal blocks are equal to random GOE matrix elements
multiplied by a coupling parameter $0\leq\lambda\leq1$. By varying
$\lambda$ between $1$ and $0$ one can change the coupling between
the diagonal blocks of the matrix. In this way a transition
between GOE case ($\lambda = 1$) and $M$-GOE case ($\lambda = 0$)
can be achieved. In our numerical study of the parametric level
dynamics of coupled M-GOE matrices we applied the Hamiltonian
model (\ref{ParametricHAmiltonian}), but instead of $H_{1}$ and
$H_{2}$ to be GOE matrices we used two $400\times 400$ coupled
M-GOE  matrices.

\begin{figure}[!]
\begin{center}
\rotatebox{270} {\includegraphics[width=0.6\textwidth,
height=0.6\textheight, keepaspectratio]{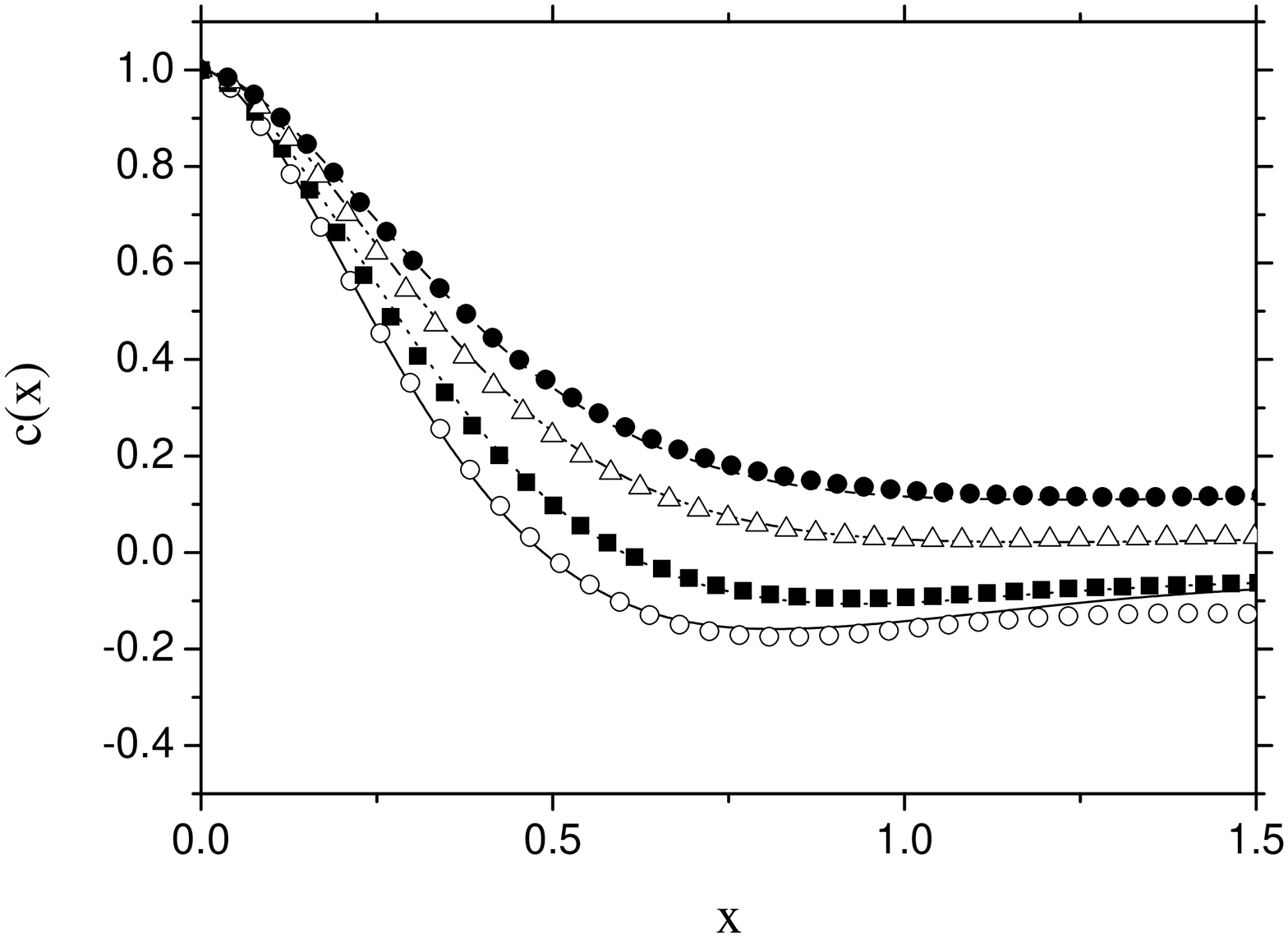}} \caption{The level
velocity autocorrelation function $c(\mathsf{x})$ for quantum
graphs with: $n=6$ (open circles), $n=7$ (full squares), $n=10$
(open triangles) and $n=30$ vertices (full circles) compared to
the results of RMT (solid line) and coupled M-GOE matrices: $M=3$,
$\lambda = 0.08$ (dotted line); $M=12$, $\lambda = 0.05$
(dash-doted line) and $M=50$, $\lambda = 0.03$ (dashed line).}
\label{Fig5}
\end{center}
\end{figure}

Let us consider the autocorrelation function $c(\mathsf{x})$ for
graphs with $n \ge 6$ vertices. Figure~\ref{Fig3} shows that the
parametric dynamics of graphs with $n=6$ is closely described by
RMT for GOE, while large discrepancies observed for  graphs with
$n>6$ suggests that their properties can be rather described by
coupled M-GOE matrices. In order to check this assumption we
fitted the values of $\lambda$ and $M$ parameters to describe
properly the behavior of the autocorrelation function
$c(\mathsf{x})$ for graphs with $n\geq7$. In Figure~\ref{Fig5} the
results obtained for quantum graphs with $n=7,10$ and $30$ are
compared with corresponding curves for coupled M-GOE matrices with
$M=3$, $\lambda=0.08$; $M=12$, $\lambda=0.05$ and $M=50$,
$\lambda=0.03$, respectively. In all cases good qualitative
agreement between the results for quantum graphs (symbols in
Figure~\ref{Fig5}) and the corresponding ones for coupled M-GOE
matrices (lines in Figure~\ref{Fig5}) is observed.

\begin{figure}[!]
\begin{center}
\rotatebox{270} {\includegraphics[width=0.5\textwidth,
height=0.6\textheight, keepaspectratio]{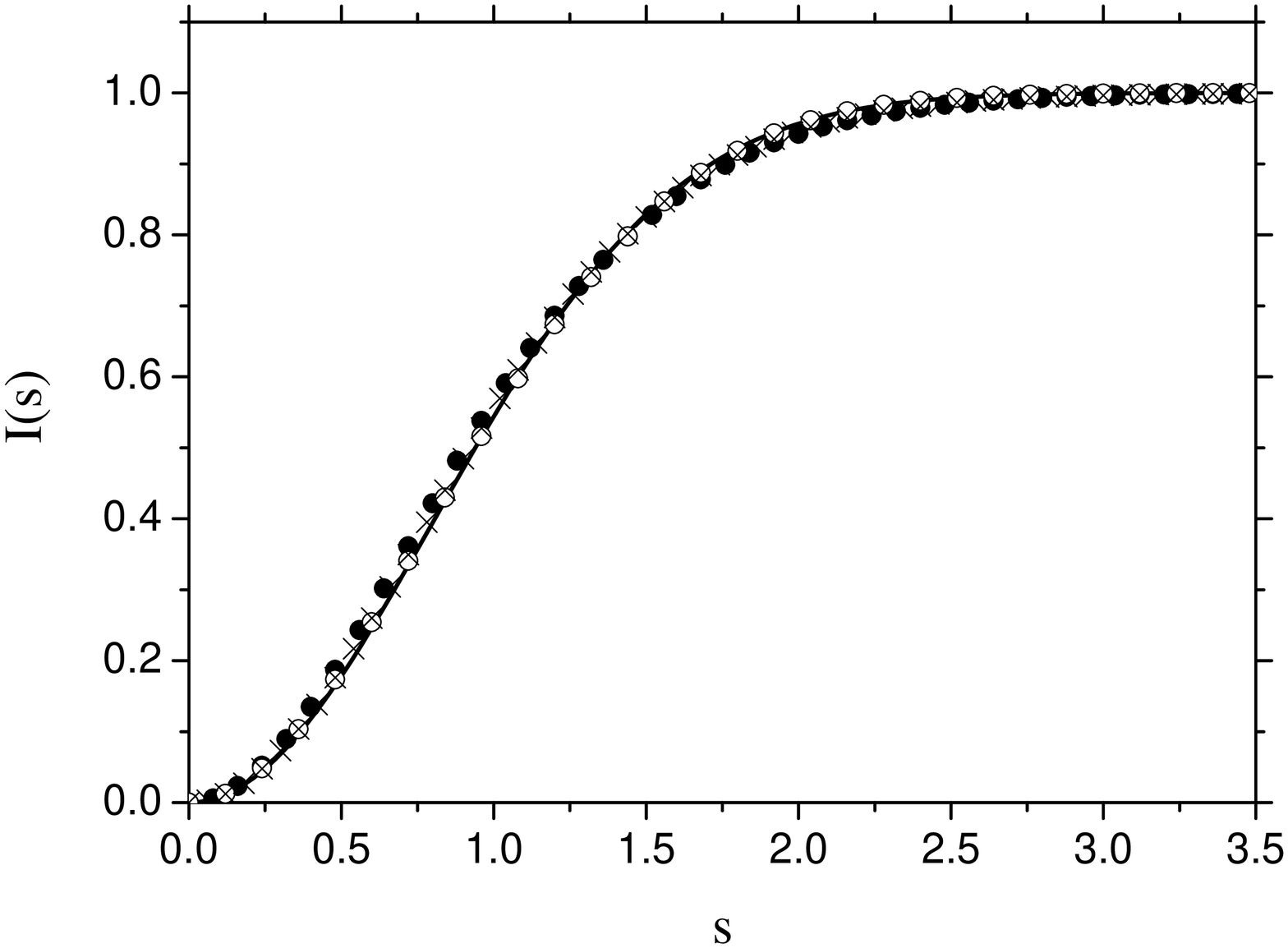}} \caption{The
integrated nearest neighbor spacing distribution (INNS) for
quantum graphs with $n=6$ (open circles), $n=30$ (full circles)
and coupled M-GOE matrices with $M=50$ and $\lambda=0.03$
(crosses) compared to the results of RMT for GOE (solid line).}
\label{Fig6}
\end{center}
\end{figure}

We also calculated the integrated nearest neighbor spacing
distribution (INNS) for quantum graphs and coupled M-GOE matrices
(see Fig.~\ref{Fig6}). We found that INNS calculated for graphs
with the size between $n=4$ and $n=30$ (in Fig.~\ref{Fig6} we show
only results for n=6 and n=30) and for coupled M-GOE matrices (in
Fig.~\ref{Fig6} we show results for matrices with $M=50$ and
$\lambda=0.03$) is in good agreement with the prediction of RMT
for GOE. This result is very interesting, while INNS and the
velocity distribution calculated for quantum graphs are in
agreement with RMT for GOE, the autocorrelation function
$c(\mathsf{x})$ departures from RMT predictions for GOE and
depends on the graph size $n$. It seems that $c(\mathsf{x})$ is
much more sensitive to non-universal features of the spectrum and
wave functions of the graphs such as eg., scars
~\cite{SchanzScars,Kaplan01} than the other considered statistics.

Another important quantity connected with parametric correlations
is the velocity autocorrelation function
$\tilde{c}(\omega,\mathsf{x})$. From the practical point of view
the definition given by the formula (\ref{Correlation}) is not
suitable because it contains delta-function $\delta(x)$. The
presence of delta-function means that one should calculate the
correlations of level velocities of energy levels separated in
energies by exactly the value of $\omega$ from one another. In
practice the number of such  levels is close to zero. That is why
we substituted infinitely narrow delta-function by the weight
function $f(x,\delta)$ with the finite width $\delta$, which
allows to define an autocorrelation function
$\tilde{c}_{\delta}(\omega,\mathsf{x})$ that is more suitable from
the practical point of view ~\cite{SimonsPRBuni}
\begin{equation}
\label{CorrelationDelta}
\tilde{c}_{\delta}(\omega,\mathsf{x})=\frac{\sum\limits_{ij}\Bigl<f\bigl(\xi_{i}(\bar{\mathsf{x}}+\mathsf{x})
-\xi_{j}(\bar{\mathsf{x}})-\omega,\sqrt{2}\delta\bigr)\frac{\partial
\xi_{i}(\bar{\mathsf{x}}+\mathsf{x})}{\partial
\bar{\mathsf{x}}}\frac{\partial
\xi_{j}(\bar{\mathsf{x}})}{\partial
\bar{\mathsf{x}}}\Bigr>}{\sum\limits_{kl}\bigl<f\bigl(\xi_{k}(\bar{\mathsf{x}}+\mathsf{x})
-\xi_{l}(\bar{\mathsf{x}})-\omega,\sqrt{2}\delta\bigr)\bigr>}\mbox{.}
\end{equation}
In our calculations we used a Gaussian weight function
$f(x,\delta)$~\cite{SimonsPRBuni}
\begin{equation}
f(x,\delta)=\frac{1}{\sqrt{2\pi}\delta}\exp\biggl(-\frac{x^{2}}{2\delta^{2}}\biggr)\mbox{.}
\end{equation}
The autocorrelation function
$\tilde{c}_{\delta}(\omega,\mathsf{x})$ defined by the formula
(\ref{CorrelationDelta}) allows to calculate correlations of level
velocities of energy levels separated mutually by $\omega\pm
\delta$.

\begin{figure}[!]
\begin{center}
\rotatebox{0} {\includegraphics[width=0.5\textwidth,
height=0.8\textheight, keepaspectratio]{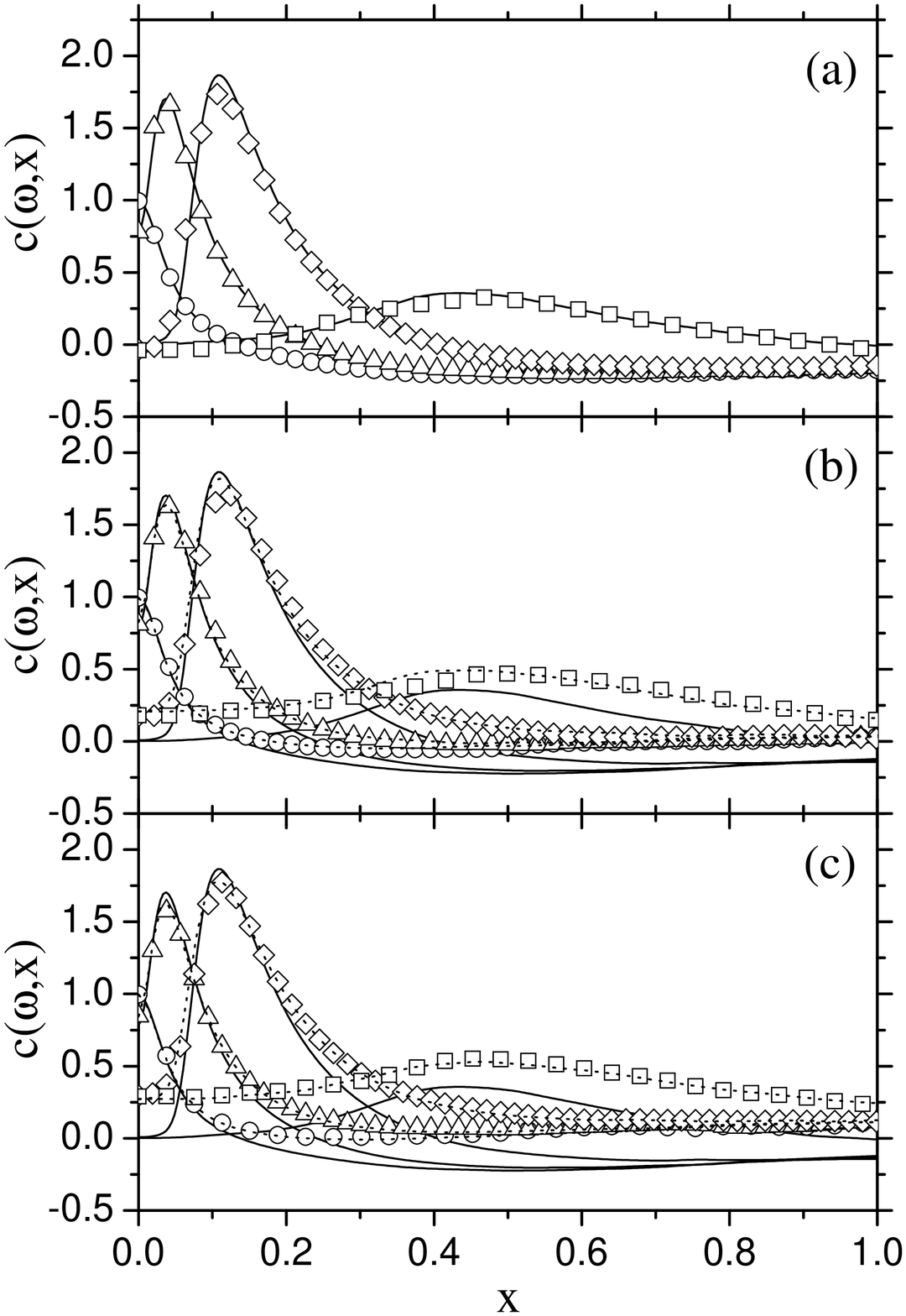}} \caption{The
velocity autocorrelation function $\tilde{c}(\omega,\mathsf{x})$
for quantum graphs with (a) $n=6$, (b) $n=10$ and (c) $n=30$
compared to the results of RMT for GOE (solid lines) and coupled
M-GOE matrices (broken lines). The autocorrelation function
$\tilde{c}(\omega,\mathsf{x})$ is calculated for four values of
the parameter $\omega$: $\omega=0$ (circles), $\omega=0.1$
(triangles), $\omega=0.25$ (diamonds), and $\omega=1$ (squares).}
\label{Fig7}
\end{center}
\end{figure}

Figures~\ref{Fig7}a, \ref{Fig7}b and \ref{Fig7}c show the
autocorrelation function of level velocities
$\tilde{c}_{\delta}(\omega,\mathsf{x})$ for quantum graphs with
$n=6$, $n=10$ and $n=30$ vertices, respectively. Calculations were
performed for four values of the parameter  $\omega=0, 0.1, 0.25$
and $1.0$. Similarly to the paper~\cite{SimonsPRBuni} the
parameter $\delta$ was chosen to be equal to $0.03$. The numerical
curves obtained for quantum graphs are compared with the
theoretical ones for  GOE in the case of graphs with $n=6$
(Fig.~\ref{Fig7}a) and for coupled M-GOE matrices for graphs with
$n=10$ and $n=30$ (Fig.~\ref{Fig7}b and Fig.~\ref{Fig7}c,
respectively). In each case the statistical averaging was
performed over $99$ graphs configurations.

We found the autocorrelation function
$\tilde{c}_{\delta}(\omega,\mathsf{x})$ calculated for graphs with
$n=6$ vertices (Fig.~\ref{Fig7}a) to be in good agreement with the
theoretical prediction for GOE (solid lines) for all the values of
the $\omega$ parameter. For quantum graphs with $n=10$ and $n=30$
vertices (Fig.~\ref{Fig7}b and Fig.~\ref{Fig7}c) the numerical
results are in agreement with the ones obtained for coupled M-GOE
matrices (broken lines in Fig.~\ref{Fig7}b and Fig.~\ref{Fig7}c)
for most values of the parameter $\mathsf{x}$. This fact confirms
that coupled M-GOE matrices can be successfully used for the
description of the parametric dynamics of quantum graphs. However,
for some values of $\mathsf{x}$ small deviations from coupled
M-GOE matrices results are observed. In Fig.~\ref{Fig7}a and
Fig.~\ref{Fig7}b the biggest deviation is observed for
$\omega=0.25$ in the vicinity of the maximum of the
autocorrelation function $\tilde{c}_{\delta}(0.25,\mathsf{x})$.

Let us consider now the level curvatures $\tilde \kappa$, the
second derivative of the unfolded energies:
\begin{equation}
\tilde \kappa_{i}=\frac{\partial^{2} \xi_{i}}{\partial
\mathsf{x}^{2}} \mbox{.}
\end{equation}

\begin{figure}[!]
\begin{center}
\rotatebox{0} {\includegraphics[width=0.5\textwidth,
height=0.8\textheight, keepaspectratio]{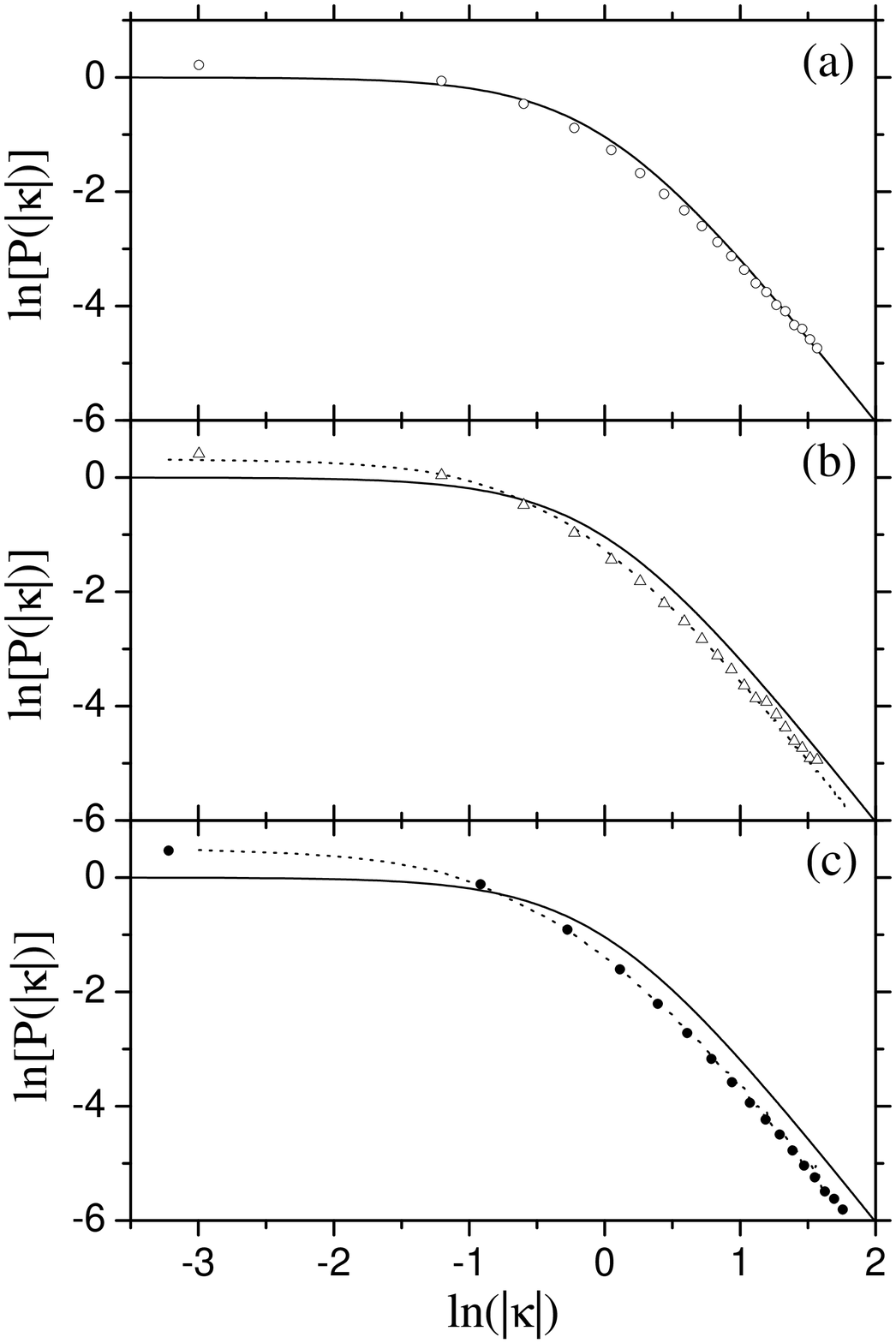}} \caption{The
parametric curvature distribution $P(\kappa)$ for quantum graphs
with (a) $n=6$ (open circles), (b) $n=10$ (open triangles) and (c)
graphs with $n=30$ vertices (full circles). The numerical results
are compared to the results of RMT for GOE (solid line) and for
coupled M-GOE matrices (broken line). Results are presented in
double logarithmic scale.} \label{Fig8}
\end{center}
\end{figure}

We rescaled the curvatures for quantum graphs using the variance
$\sigma_{\mathsf{v}}$:
\begin{equation}
\kappa_{i}=\frac{\tilde \kappa_{i}} {\sigma_{\mathsf{v}}} \mbox{.}
\end{equation}
The distributions of level curvatures are presented in
Fig.~\ref{Fig8}a, Fig.~\ref{Fig8}b and Fig.~\ref{Fig8}c for graphs
with $n=6$, $n=10$ and $n=30$, respectively. It is visible that
for all of the cases there is no agreement of the obtained results
with the GOE prediction (solid line). Only for the graphs with
$n=6$ vertices the tail of the distribution can be described
correctly by the GOE prediction. For the other graphs the level
curvature distributions departure from the GOE one. Therefore, in
the case of graphs with $n=10$ and $n=30$ we compare the level
curvature distribution with the results obtained for coupled M-GOE
matrices (broken line). For the graphs with $n=10$ vertices
(Fig.~\ref{Fig8}b) small deviations from the results for coupled
M-GOE matrices are observed only for small values of level
curvatures. In the case of graphs with $n=30$ (Fig.~\ref{Fig8}c)
the curvature distribution calculated for coupled M-GOE matrices
(broken line) follows the results obtained for the graphs for all
of the range of the parameter $\kappa$.

One of the characteristic features of level dynamics of quantum
chaotic systems are avoided crossings which are often observed
between the neighboring levels when an external parameter $X$ is
changed (see Fig.~\ref{Fig1} and Fig.~\ref{Fig2}). We define the
avoided crossing distance $C$ as a local minimum of the distance
between two neighboring eigenvalues. The distribution of avoided
crossings $P(c)$ can be used to distinguish between chaotic
systems which belong to different universality classes. The
analytical formulas for the distributions of avoided crossings
$P(c)$ for GOE  and GUE  were derived by Zakrzewski and
Ku\'s~\cite{ZakrzewskiKusAcross}. For GOE-systems the distribution
of avoided crossings  $P(c)$ is the following:
\begin{equation}
\label{AcrossGOE}
P(c)=\frac{2}{\pi}\exp\biggl(-\frac{c^{2}}{\pi}\biggr)\mbox{.}
\end{equation}

To calculate the local minimum $C$ of the avoided crossing
distance we used the hyperbolic formula~\cite{ZakrzewskiAcross}:
\begin{equation}
\label{HyperbolicApproach}
C=\sqrt{d_{2}^{2}-\frac{(d_{3}^{2}-d_{1}^{2})^{2}}{8(d_{3}^{2}+d_{1}^{2}-2d_{2}^{2})}}\mbox{,}
\end{equation}
where $d_{i}$, $i=1,2,3$ are  three consecutively calculated
distances between the adjacent levels. Then avoided crossing
distances $C$ obtained in this way are rescaled using the formula:
\begin{equation}
c=\frac{C}{\langle C\rangle}\mbox{,}
\end{equation}
where $\langle C\rangle$ is the mean value of avoided crossing
distance.

\begin{figure}[!]
\begin{center}
\rotatebox{270} {\includegraphics[width=0.5\textwidth,
height=0.6\textheight, keepaspectratio]{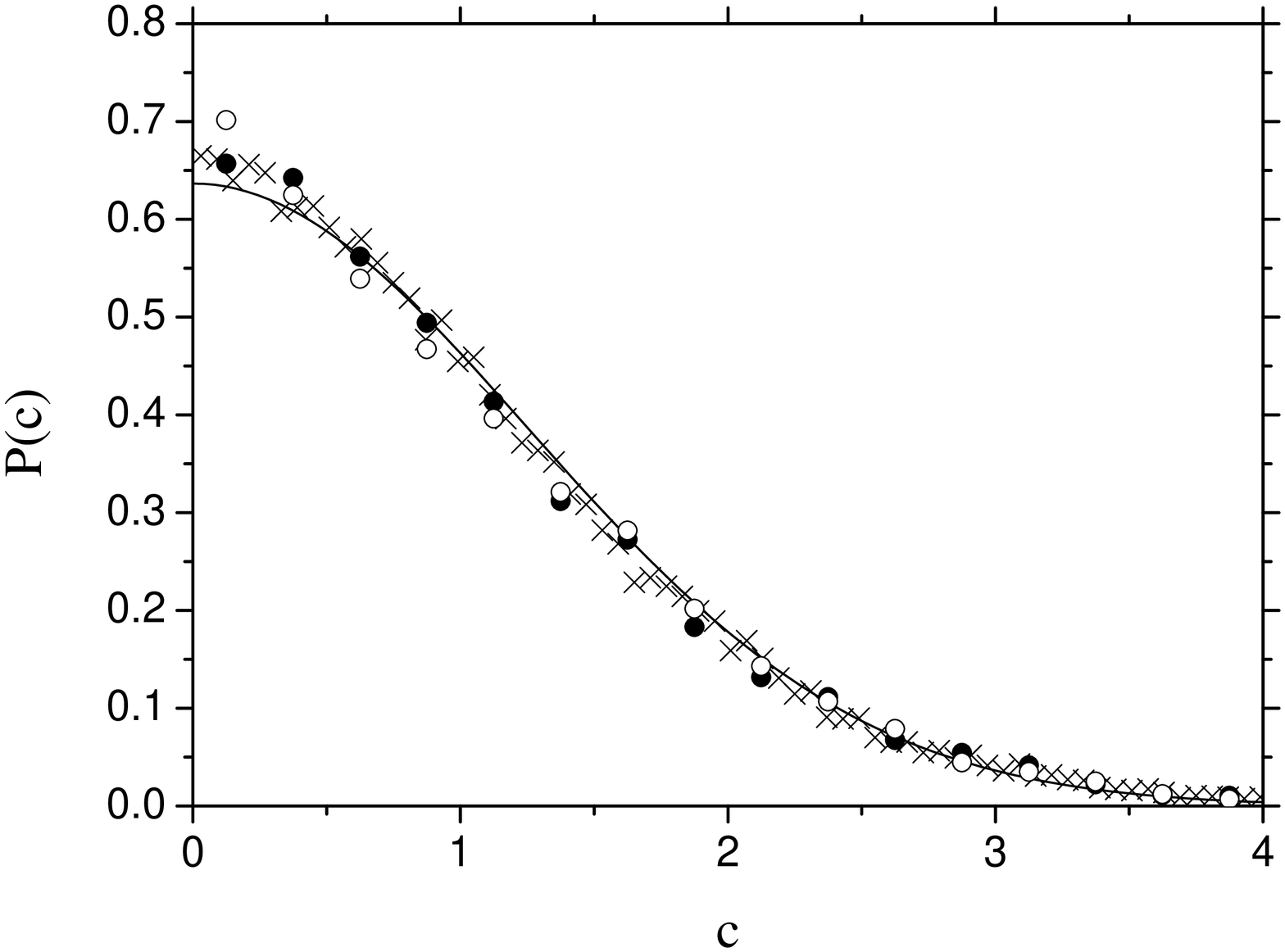}} \caption{The
distribution of avoided crossings $P(c)$ for quantum graphs with
$n=6$ (open circles), $n=30$ (full circles) and coupled M-GOE
matrices with $M=50$ and $\lambda=0.03$ (crosses) compared to the
results of RMT for GOE (solid line).} \label{Fig9}
\end{center}
\end{figure}

Figure~\ref{Fig9} shows the distribution of avoided crossings
$P(c)$ for quantum graphs with  $n=6$ (open circles), $n=30$ (full
circles) and coupled M-GOE matrices with $M=50$ blocks (crosses)
compared to the results of RMT for GOE (solid line). The numerical
results for graphs were obtained as a result of an averaging over
$99$ graph configurations. The distributions of avoided crossings
$P(c)$ obtained for graphs show an excess of small values compared
to the theoretical prediction for GOE (solid line). However, it is
worth noting that in the range of $0.8<c<1.5$ the results obtained
for the graphs with $n=6$ vertices are slightly smaller than the RMT
prediction for GOE. The results obtained for coupled M-GOE
matrices show also an excess of small values comparable to the GOE
prediction. However, for $c>0.5$ M-GOE results are close to the ones
obtained for the RMT prediction for GOE as well as for the graphs with $n=30$ vertices.

In summary, we investigated numerically eigenvalue dynamics of
fully connected irregular quantum graphs with Neumann boundary
conditions of different size $n=4-30$. We showed that there are
some spectral statistics such as the integrated nearest neighbor
spacing distribution, the parametric velocity distribution $P(v)$
and the distribution of avoided crossings $P(c)$ that show no or
weak sensitivity to the system size. These statistics show
GOE-like behavior. There are also some other statistics, e.g., the
second order level velocity autocorrelation functions
$c(\mathsf{x})$ and $\tilde{c}(\omega,\mathsf{x})$ and the
parametric curvature distribution $P(\kappa )$ that for larger
graphs ($n > 6$) show deviations from the predictions of RMT for
GOE. In all these cases the obtained results are much better
described by the model of coupled M-GOE matrices. The agreement of
the obtained results with the coupled M-GOE matrices predictions
supported by the results of the inverse participation ratio and
the distributions $P(\Psi(x))$ suggest the existence of
non-ergodic structures  of graphs' wave functions connected with
localization which cause the quantum dynamical disintegration of
large graphs into the union of weakly interacting smaller graphs.

It is worth pointing out that the agreement of parametric spectral
statistics of quantum graphs with the predictions of RMT could be
restored by making quantum graphs "more ergodic". It could be done
by randomizing the boundary conditions at the vertices. For
example, one can replace vertex-scattering matrices by random
matrices taken from Circular Orthogonal Ensemble (COE)
\cite{KottosCOE}. Our preliminary results, which are not presented
here, suggest that in this case parametric spectral statistics
show good agreement with RMT predictions for GOE both for smaller
and larger graphs. Moreover, this agreement becomes better for
larger graphs.

Acknowledgments. We would like to thank Szymon Bauch for valuable
comments. This work was supported by the Ministry of Science and
Higher Education grant No. N202 099 31/0746. One of the authors,
PS, was also supported by  the Ministry of Education, Youth and
Sports of the Czech Republic within the project LC06002.

\end{document}